\documentclass[aps,prb,twocolumn,showpacs,superscriptaddress]{revtex4-1}

\usepackage{graphicx}
\usepackage{dcolumn}
\usepackage{color}
\usepackage{array,multirow,makecell}
\usepackage{bm}
\usepackage{nicefrac}

\begin{document}
\title{High-field moment polarization in the itinerant ferromagnet URhSi}

\author{W. Knafo}
\affiliation{Laboratoire National des Champs Magn\'etiques Intenses, UPR 3228, CNRS-UPS-INSA-UGA, 143 Avenue de Rangueil, 31400 Toulouse, France}

\author{T.D. Matsuda}
\affiliation{Department of Physics, Tokyo Metropolitan University, Hachioji, Tokyo 192-0397, Japan}

\author{F. Hardy}
\affiliation{Institut f\"{u}r Festk\"{o}rperphysik, Karlsruher Institut f\"{u}r Technologie, 76021 Karlsruhe, Germany.}

\author{D. Aoki}
\affiliation{Institute for Materials Research, Tohoku University, Ibaraki 311-1313, Japan}

\author{J. Flouquet}
\affiliation{Universit\'{e} Grenoble Alpes and CEA, INAC-PHELIQS, F-38000 Grenoble, France}


\date{\today}

\begin{abstract}
We report a high-magnetic-field study of the itinerant ferromagnet URhSi. Magnetization and electrical resistivity were measured under magnetic fields $\mu_0H$ up to 58~T applied along the directions $\mathbf{a}$, $\mathbf{b}$, and $\mathbf{c}$ of the orthorhombic structure and temperatures $T$ ranging from 1.5 to 50 K. For $\mathbf{H}\parallel\mathbf{b}$, pseudo-metamagnetism at $\mu_0H_m\simeq30-40$~T is associated with a broad step in the magnetization and a maximum in the resistivity. The properties of URhSi are discussed and compared with those of the isostructural superconducting ferromagnets URhGe and UCoGe and of the superconducting paramagnet UTe$_2$.

\end{abstract}

\maketitle

\section{Introduction}

While superconductivity sets in at the antiferromagnetic instability of a large number of materials including heavy-$f$-electron,\cite{Pfleiderer2009} Fe-based,\cite{Johnston2010} and organic\cite{Lefebvre2000} systems, its occurrence close to a ferromagnetic instability was reported only in a few U-based heavy-electron systems.\cite{Aoki2019} Superconductivity develops in the ferromagnets URhGe [\onlinecite{Aoki2001}] and UCoGe [\onlinecite{Huy2007}] at ambient pressure and UGe$_2$ [\onlinecite{Saxena2000}] and UIr [\onlinecite{Akazawa2004}] under pressure. Recently, the paramagnet UTe$_2$, which is suspected to lie in the vicinity of a ferromagnetic instability, was found to become superconducting at ambient pressure too.\cite{Ran2019,Aoki2019b} Under a magnetic field applied along a hard magnetic axis ($\mathbf{H}\parallel\mathbf{b}$), a reentrance of superconductivity or an anomalous S-shape of the superconducting field have been reported in URhGe [\onlinecite{Levy2005}], UCoGe ($\mathbf{H}\parallel\mathbf{b}$) [\onlinecite{Aoki2009}], and UTe$_2$ [\onlinecite{Ran2019},\onlinecite{Knebel2019},\onlinecite{Ran2019b}] at ambient pressure. In UGe$_2$ under pressure, a S-shape of the superconducting field was also reported in a field applied along the easy magnetic axis ($\mathbf{H}\parallel\mathbf{a}$) [\onlinecite{Sheikin2001}].

A challenge in these U-based magnets is to understand the relationship between field-induced superconductivity and metamagnetism, in which changes of the magnetic fluctuations possibly couple to a Fermi surface reconstruction. In URhGe, a metamagnetic transition, evidenced as a field-induced first-order step in the magnetization at $\mu_0H_m=12$~T, and associated with a reorientation of the moments (rotating from the $\mathbf{c}$ to the $\mathbf{b}$ axes), coincides with the field-reentrance of superconductivity.\cite{Levy2005} In this material, hydrostatic pressure leads to the increase of $H_m$ and the disappearance of reentrant superconductivity,\cite{Miyake2009} while uniaxial pressure applied along $\mathbf{b}$ reduces $H_m$ and boosts the superconducting temperature.\cite{Braithwaite2018} In UTe$_2$, field-induced superconductivity also develops in the vicinity of a metamagnetic transition at $\mu_0H_m\simeq35$~T [\onlinecite{Knebel2019},\onlinecite{Knafo2019},\onlinecite{Miyake2019},\onlinecite{Ran2019b}]. The situation in UCoGe is more subtle, since a fall of $T_C$ observed by magneto-transport at $\mu_0H^*=15$~T coincides with a S-shape of the superconducting borderline $H_{sc}$ [\onlinecite{Aoki2009}], while a pseudo-metamagnetic crossover is observed at a much higher field $\mu_0H_m\simeq50$~T [\onlinecite{Knafo2012}].  To understand the interplay between magnetism and superconductivity in U-based ferromagnets and nearly-ferromagnets, a target is, thus, to characterize their high-field properties, and in particular, their high-field moment polarization processes.

We present here a high-magnetic field study of the itinerant ferromagnet URhSi, which is isostructural to the ferromagnetic superconductors URhGe and UCoGe (Pnma orthorhombic structure), but where superconductivity has not been observed yet \cite{Tran1998,Prokes2003,Tateiwa2019} (see Table \ref{table}). The Curie temperature $T_C=10.5$~K of URhSi is quite comparable to that ($T_C=9.5$~K) of URhGe. For the three materials, magnetic susceptibility measurements indicate similar Ising anisotropies, where $\mathbf{c}$ is the easy magnetic axis, $\mathbf{a}$ is the hardest magnetic axis, and $\mathbf{b}$ is an intermediate hard magnetic axis.\cite{Hardy2011,Knafo2012,Matsuda2019} In this study, we have studied the field-induced properties of URhSi in fields applied along the three crystallographic directions $\mathbf{a}$, $\mathbf{b}$, and $\mathbf{c}$. In a magnetic field applied along $\mathbf{b}$, we observe and characterize a broad pseudo-metamagnetic crossover in URhSi. In the discussion, comparison is made with similar metamagnetic phenomena observed in the superconducting ferromagnets URhGe, UCoGe and paramagnet UTe$_2$.

\begin{table}[t]
\caption{Curie temperature $T_C$, spontaneous moment $M_s$, superconductive temperature $T_{SC}$ at $H=0$ and, for $\mathbf{H}\parallel\mathbf{b}$, temperature $T_{\chi}^{max}$ at the maximum of the magnetic susceptibility, metamagnetic field $H_m$, and maximum $T_{RSC}^{max}$ of the reentrant superconductive temperature, in URhGe, UCoGe, URhSi, and UTe$_2$.}
\begin{ruledtabular}
\begin{tabular}{lcccc}
($p=1$~bar)&URhGe&UCoGe&URhSi& UTe$_2$\\
\hline
$T_C$ (K)&9.5&3&10.5&-\\
$M_s$ ($\mu_B$)&0.4&0.05&0.5&-\\
$T_{SC}$ ($H=0$) (K)&0.25&0.6&-&1.6\\
$T_{\chi}^{max}$ [$\mathbf{H}\parallel\mathbf{b}$] (K)&9.5&37.5&-&35\\
$\mu_0H_{m}$ [$\mathbf{H}\parallel\mathbf{b}$] (T)&12&50&30-40&35\\
$T_{RSC}^{max}$ [$\mathbf{H}\parallel\mathbf{b}$] (K)&0.6&- $^\dag$&-&1\\
\hline
Refs.&[\onlinecite{Aoki2001},\onlinecite{Knafo2012}]&[\onlinecite{Huy2007},\onlinecite{Aoki2009},\onlinecite{Knafo2012}]&[\onlinecite{Tran1998},\onlinecite{Prokes2003}]&[\onlinecite{Ran2019},\onlinecite{Knebel2019},\onlinecite{Knafo2019},\onlinecite{Miyake2019}]\\
\end{tabular}
\end{ruledtabular}
 $^\dag$ In UCoGe, a $S$-shape of $H_{c,2}$ is observed at 10-15~T [\onlinecite{Aoki2009}].
\label{table}
\normalsize
\end{table}

\section{Experimental techniques}

Single crystals were grown by the Czochralski-pulling method in a tetra-arc furnace with a stoichiometry composition under argon-gas atmosphere (see Ref. [\onlinecite{Matsuda2019}] for more details). Crystal orientation was checked by x-ray diffraction. Pulsed magnetic fields were generated using standard 60-T magnets at the LNCMI-Toulouse pulsed field facility (France). Magnetization was measured using compensated coils on a $1.3\times1\times0.9$~mm$^3$ sample under pulsed magnetic fields up to 53~T. Pulsed-field magnetization data were adjusted (to eliminate an offset) to low-field magnetization data measured up to 5~T using a commercial 'MPMS' magnetometer from Quantum Design. Electrical resistivity under pulsed magnetic fields up to 58~T was measured by the four-contacts technique, and extracted using a numerical lock-in, on a $1\times0.3\times0.1$~mm$^3$ sample with a current of 10 mA, with a frequency $\simeq40-70$~kHz, applied along the $\mathbf{a}$ direction.

\section{High magnetic-field properties}

\begin{figure}[t]
\includegraphics[width=1\columnwidth]{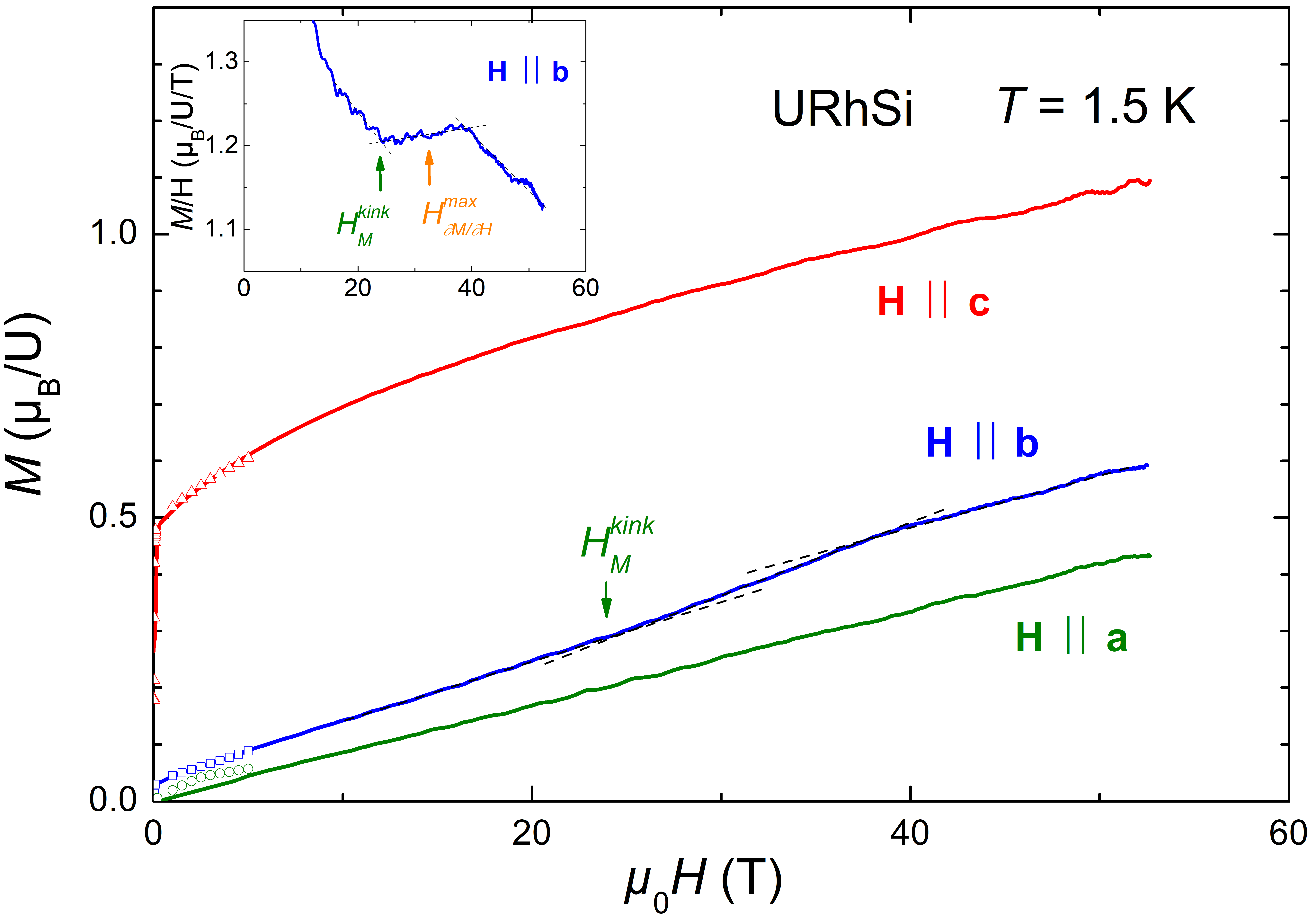}
\caption{Magnetization $M$ versus magnetic field $H$ of URhSi at the temperature $T=1.5$~K, for $\mathbf{H}\parallel\mathbf{a},\mathbf{b},\mathbf{c}$. Open-symbol data were measured under steady fields, full-line data were measured under pulsed fields. Some oscillations in the data correspond to non-physical noise. The inset shows a zoom on $M/H$ versus $H$ data for $\mathbf{H}\parallel\mathbf{b}$.}
\label{Fig1}
\end{figure}

Figure \ref{Fig1} presents the low-temperature ($T=1.5$~K) magnetization of URhSi in a magnetic field applied along its three main crystallographic directions $\mathbf{a}$, $\mathbf{b}$, and $\mathbf{c}$. Similarly to the recent report on URhSi in Ref. [\onlinecite{Tateiwa2019}], and to the cases of URhGe [\onlinecite{Hardy2011}] and UCoGe [\onlinecite{Knafo2012}], $\mathbf{a}$ is found to be the hardest magnetic axis, while $\mathbf{c}$ is the easy magnetic axis. However, a different magnetic anisotropy, with $\mathbf{b}$ being the hardest magnetic axis instead of $\mathbf{a}$, was reported from magnetization data on URhSi in Refs. [\onlinecite{Prokes2003},\onlinecite{Honda2003}]. A sensibility of the physical properties to sample mis-orientation combined with crystal mosaicity may explain the differences between the different sets of data. In a field $\mathbf{H}$ applied along $\mathbf{c}$, the spontaneous magnetization $M_s=0.5$~$\mu_B/U$ is reached at very low fields, indicating the alignment of the magnetic domains; beyond the domains alignment, the magnetization $M$ continues to increase with $H$, reaching $1.1$~$\mu_B/U$ at $\mu_0H=53$~T, and is probably associated with a field-induced quenching of magnetic fluctuations. In a magnetic field applied along $\mathbf{a}$, an almost linear magnetization is observed in the whole field range, with a slope similar to that of $M(H)$ in high fields along $\mathbf{c}$. In a field applied along the intermediate magnetic axis $\mathbf{b}$, a broad crossover can be seen in the $M(H)$ data, leading to a kink of $M/H$  at $\mu_0H_M^{kink}=24$~T and to a maximum of slope at $\mu_0H_{\partial M/\partial H}^{max}=32$~T (at $T=1.5$~K). This anomaly is identified as the signature of a pseudo-metamagnetic crossover, i.e., a similar but broader phenomenon than a first-order metamagnetic transition.

\begin{figure}[t]
\includegraphics[width=1\columnwidth]{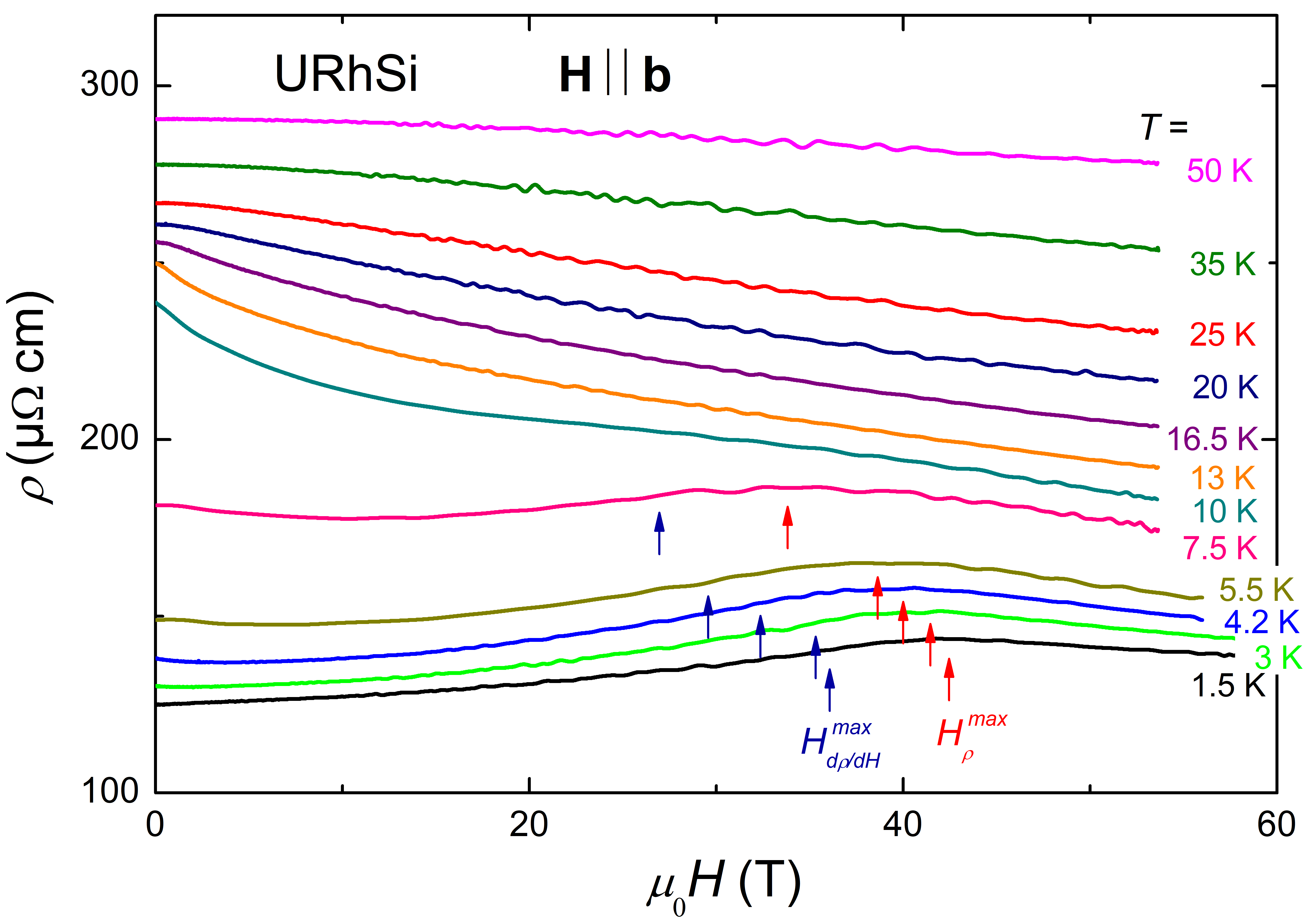}
\caption{Electrical resistivity $\rho$ versus magnetic field $H$, at different temperatures $T$ from 1.5 to 50~K, for $\mathbf{H}\parallel\mathbf{b}$.}
\label{Fig2}
\end{figure}

\begin{figure}
\includegraphics[width=1\columnwidth]{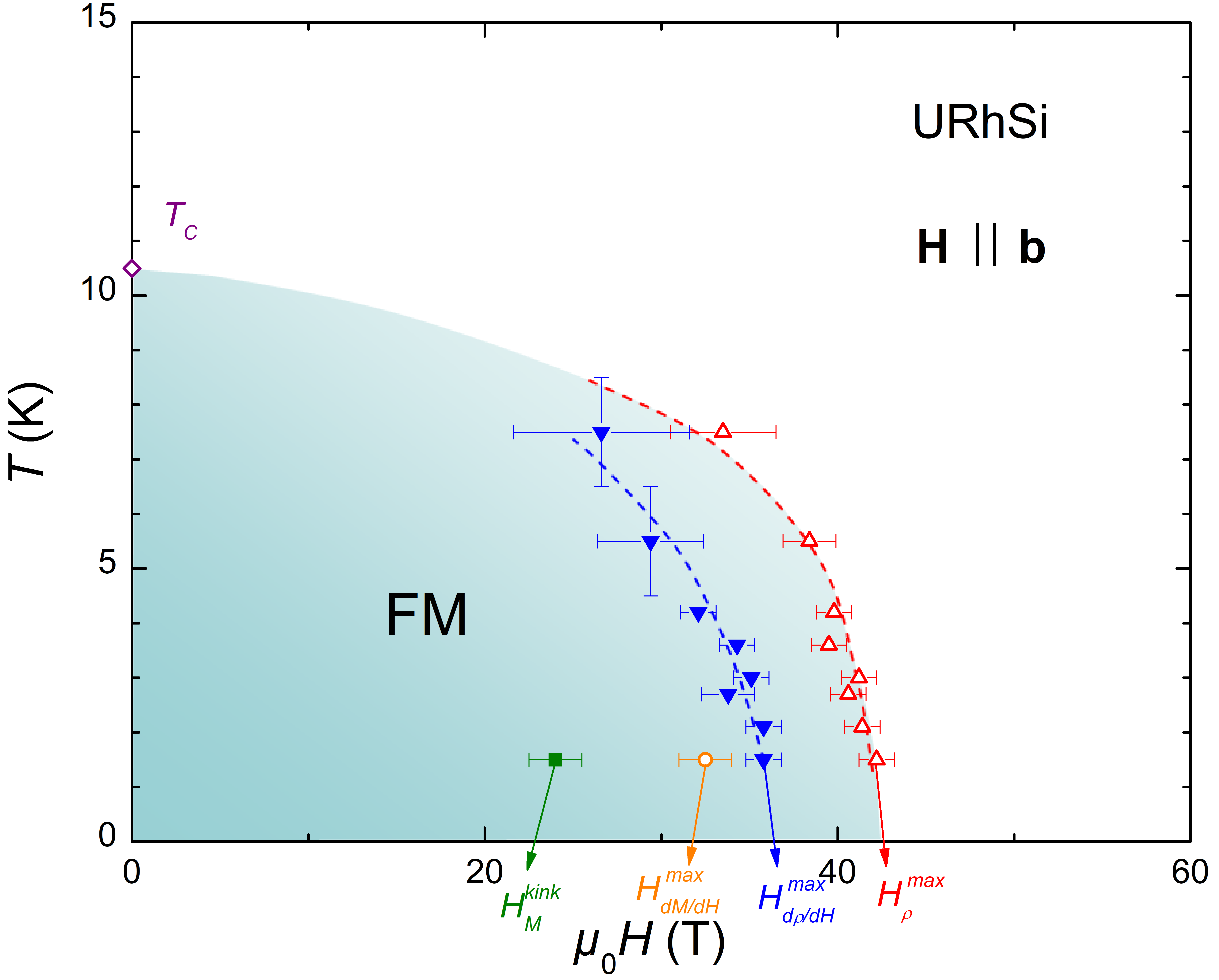}
\caption{Magnetic-field-temperature phase diagram of URhSi for $\mathbf{H}\parallel\mathbf{b}$.}
\label{Fig3}
\end{figure}

Figure \ref{Fig2} shows the resistivity $\rho$ versus magnetic field data measured for $\mathbf{H}\parallel\mathbf{b}$, where a broad maximum is observed at $\mu_0H_\rho^{max}$, which equals 42~T at $T=1.5$~K. A maximum of the slope of $\rho(H)$ at $H_{\partial\rho/\partial H}^{max}$ precedes the maximum of $\rho(H)$. The fields $H_\rho^{max}$ and $H_{\partial\rho/\partial H}^{max}$ both decrease when the temperature is raised, before disappearing at temperatures $T>T_C=10.5$~K. $H_\rho^{max}$ and $H_{\partial\rho/\partial H}^{max}$, as well as $H_M^{kink}$ and $H_{\partial M/\partial H}^{max}$, were defined using different criteria, but they all characterize the same broad moment polarization process. These different field and temperature scales are summarized in the magnetic phase diagram plotted in Figure \ref{Fig3}. In the following, we will associate the 'mean' characteristic field $\mu_0H_m=30-40$~T to the pseudo-metamagnetic crossover occurring in URhSi for $\mathbf{H}\parallel\mathbf{b}$.

\begin{figure}[t]
\includegraphics[width=1\columnwidth]{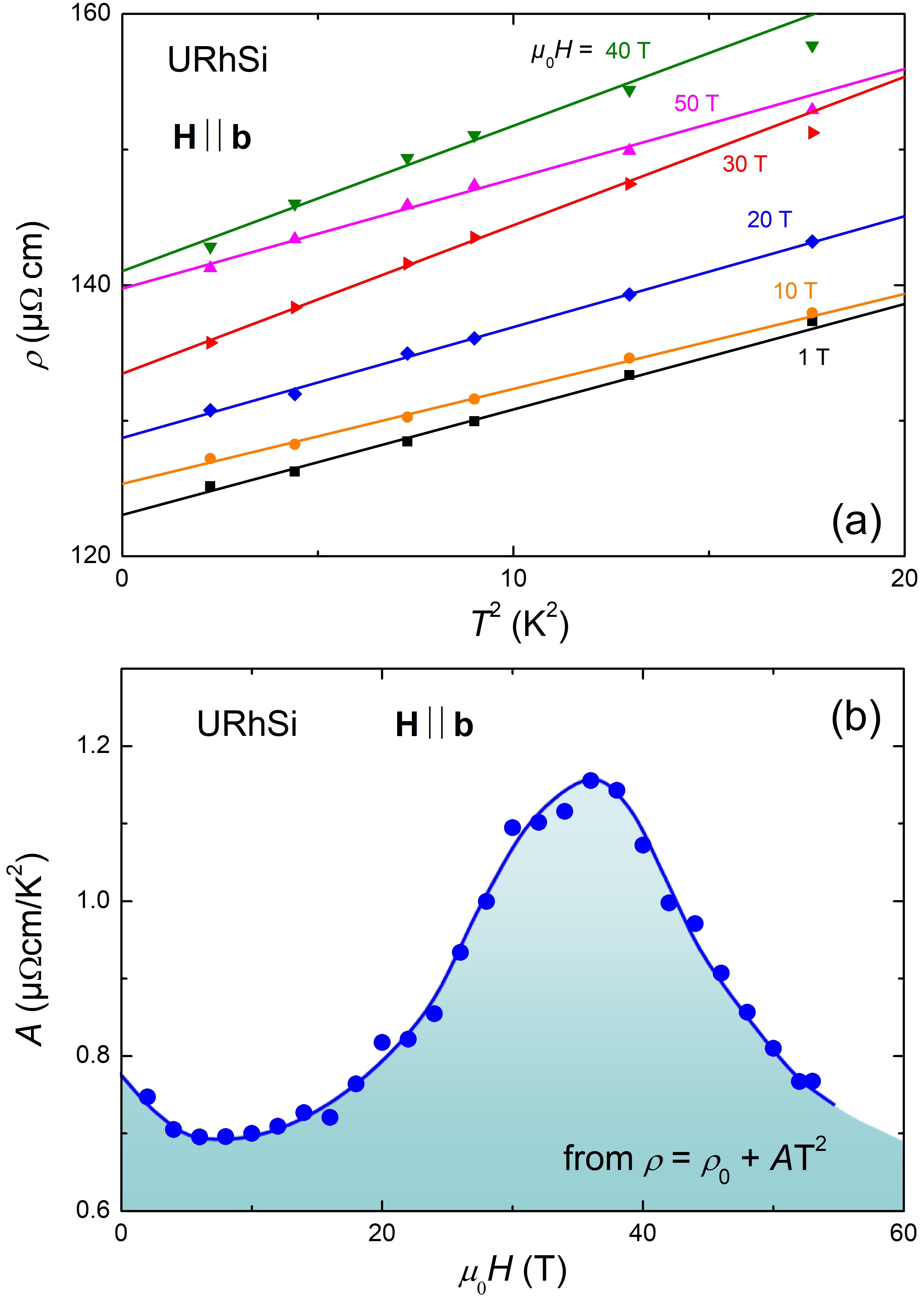}
\caption{(a) Electrical resistivity $\rho$ versus the square of temperature $T^2$, at different  magnetic field $\mu_0H$ from 1 to 50~T, for $\mathbf{H}\parallel\mathbf{b}$. (b) Magnetic field variation of the quadratic coefficient $A$ of the resistivity.}
\label{Fig4}
\end{figure}

From $\rho$ versus $T^2$ plots extracted at different magnetic fields and shown in Figure \ref{Fig4}(a), a Fermi-liquid-like fit $\rho=\rho_0+AT^2$ to the data under temperatures $T\leq4$~K permits to extract the quadratic coefficient $A$. Its magnetic-field-variation is shown in Figure \ref{Fig4}(b). After a small decrease in fields up to 5~T, the coefficient $A$ increases and passes through a maximum at $\simeq38$~T, before decreasing in higher fields. Within a crude Fermi-liquid picture assuming a Kadowaki-Woods behavior, which is often followed in heavy-fermion compounds, $A$ is proportional to the square $m^{*2}$ of the effective mass\cite{Kadowaki1986} and is driven by electronic correlations associated with quantum magnetic fluctuations (see the Ce$_{1-x}$La$_{x}$Ru$_2$Si$_2$ and URhGe cases [\onlinecite{Knafo2009},\onlinecite{Tokunaga2015},\onlinecite{Miyake2008},\onlinecite{Hardy2011}]). In URhSi, the broad step in $M(H)$ and maximum in $\rho(H)$ coincide with the maximum in $A(H)$ at $H_m$.

\begin{figure}
\includegraphics[width=1\columnwidth]{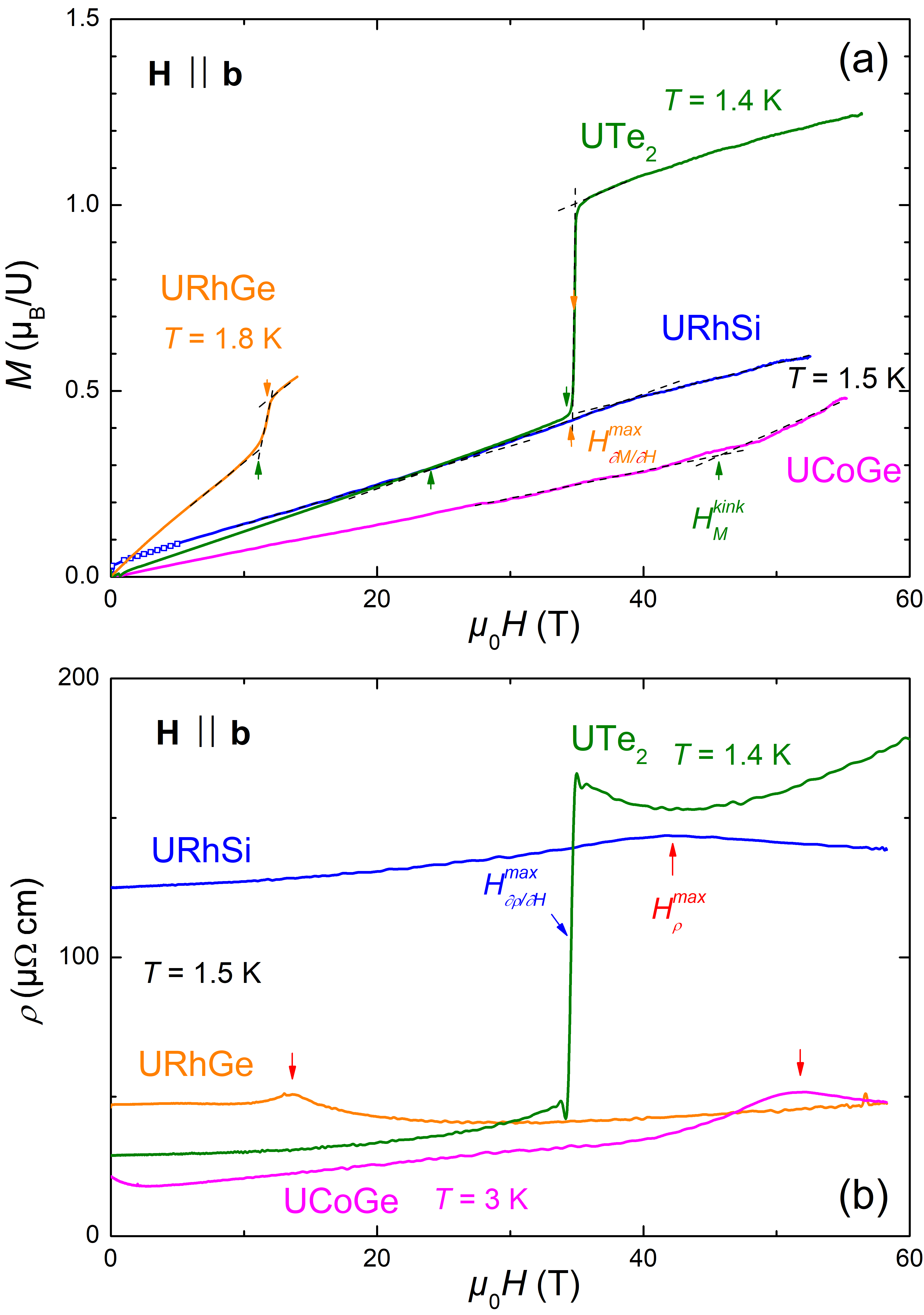}
\caption{(a) Magnetization $M$ versus magnetic field $H$ of URhSi and UCoGe at $T=1.5$~K, of URhGe at $T=1.8$~K, and of UTe$_2$ at $T=1.4$~K. (b) Electrical resistivity $\rho$ versus magnetic field $H$ of URhSi, URhGe at $T=1.5$~K, of UTe$_2$ at $T=1.4$~K and of UCoGe at $T=3$~K. (data on UCoGe, URhGe, and UTe$_2$ were taken from Refs. [\onlinecite{Knafo2012},\onlinecite{Miyake2008}])}
\label{Fig5}
\end{figure}

\begin{figure}
\includegraphics[width=1\columnwidth]{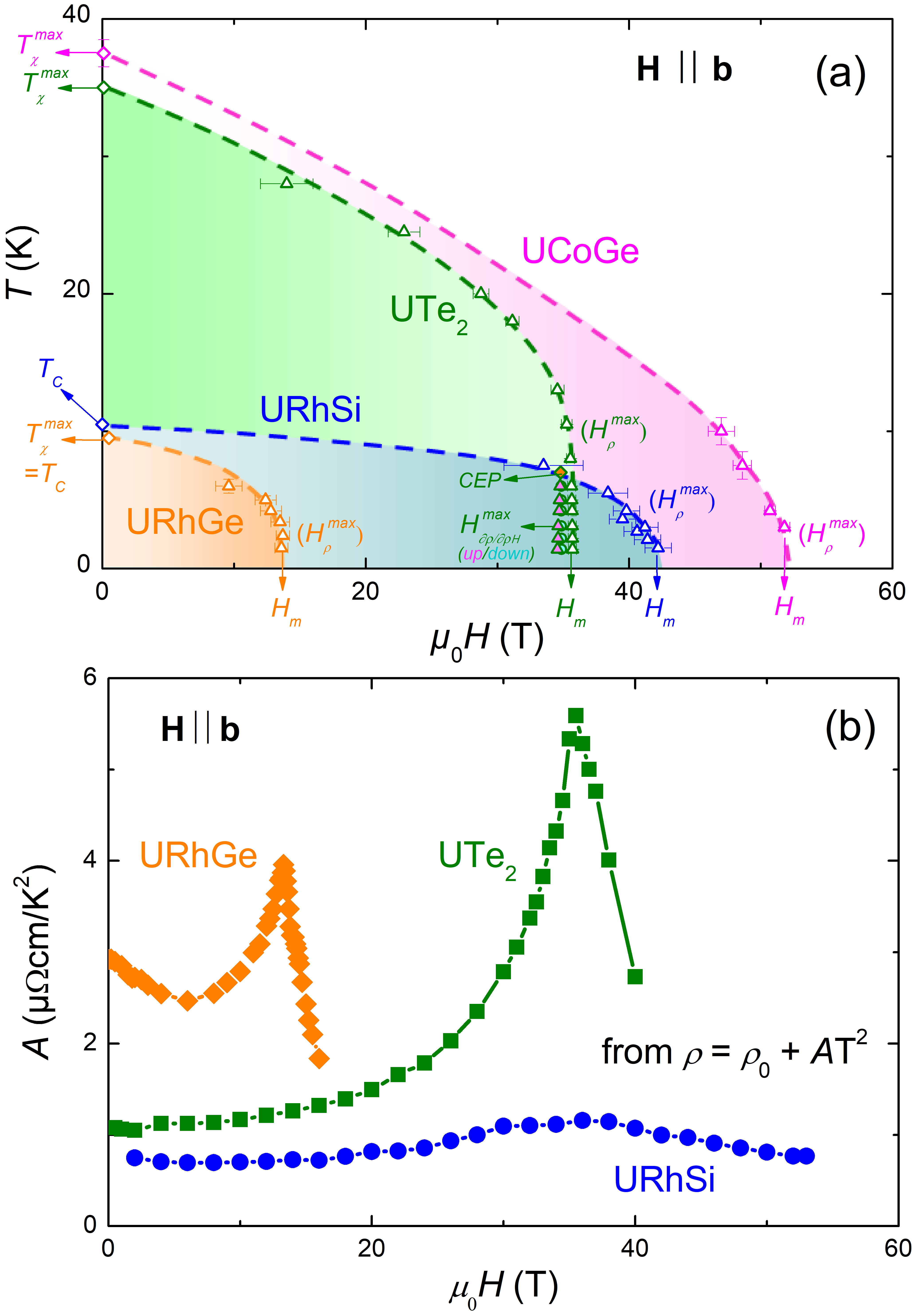}
\caption{(a) Magnetic-field-temperature phase diagram of URhSi, URhGe, UCoGe, and UTe$_2$ for $\mathbf{H}\parallel\mathbf{b}$, (b) Quadratic coefficient $A$ of the electrical resistivity versus magnetic field of URhSi, URhGe, and UTe$_2$ for $\mathbf{H}\parallel\mathbf{b}$. (data on UCoGe, URhGe, and UTe$_2$ were taken from Refs. [\onlinecite{Miyake2019},\onlinecite{Knafo2012}])}
\label{Fig6}
\end{figure}

\section{Discussion}

We compare here the high-magnetic-field properties of URhSi with those of the isostructural itinerant ferromagnets URhGe and UCoGe, and of the paramagnet UTe$_2$.

Although they all have the same Ising-type anisotropy characterized by $M_c>M_b>M_a$, the three materials URhSi, URhGe, and UCoGe present different initial slopes $(\partial M/\partial H)_i$ hierarchies in their low-temperature magnetization. While $(\partial M/\partial H)_b>(\partial M/\partial H)_c>(\partial M/\partial H)_a$ is found in URhGe [\onlinecite{Hardy2011}], we observe here the hierarchy $(\partial M/\partial H)_c>(\partial M/\partial H)_b>(\partial M/\partial H)_a$ in URhSi, which is similar to that observed in UCoGe.\cite{Knafo2012} These slopes are related to the field-quenching of the magnetic fluctuations and indicate that the magnetic fluctuations continue to subsist in high magnetic field, where a large polarization is achieved.  However, little is known about the magnetic fluctuations of these systems in high magnetic field.  At zero field, ferromagnetic fluctuations were observed in the ferromagnets UGe$_2$ and UCoGe and their magnetic anisotropy was suggested to be an important parameter for the appearance of superconductivity \cite{Huxley2003,Haslbeck2019,Stock2011}.

A key action of a magnetic field applied along the hard magnetic axis $\mathbf{b}$ is, by reducing the Curie temperature $T_C$, to drive to a ferromagnetic quantum instability. Remarkably, metamagnetism or pseudo-metamagnetism occurs in the four compounds URhSi, URhGe, UCoGe, and UTe$_2$ in a magnetic field applied along their hard magnetic axis ($\mathbf{H}\parallel\mathbf{b}$) and its onset coincides with a low-temperature magnetization reaching $M\simeq0.3-0.4$~$\mu_B$/U [see Figure \ref{Fig5}(a)]. While a sharp step-like anomaly is observed at $\mu_0H_m\simeq35$~T in UTe$_2$ and at $\mu_0H_m\simeq12$~T in URhGe, a broad anomaly is reported at $\mu_0H_m\simeq30-40$~T in URhSi. In UCoGe, due to higher field scales, only the kink at the onset of the step is observed, and the transition to the high-field regime is not completed at 53~T.
Figure \ref{Fig5}(b) shows the resistivity versus field of the four materials in a magnetic field $\mathbf{H}\parallel\mathbf{b}$, at the temperatures $T=1.5$~K for URhGe and URhSi, $T=1.4$~K for UTe$_2$, and $T=3$~K for UCoGe. Here also, the maximum in $\rho(H)$ delimiting the high-field regime is much less marked in URhSi than in UTe$_2$ and URhGe, UCoGe being in an intermediate situation. The large low-temperature and zero-field resistivity [in relation with the small residual resistivity ratio $\rho(300\textrm{K})/\rho(T\rightarrow0)\simeq2.5$] indicates the low quality of our URhSi single crystals, in comparison with the URhGe, UTe$_2$, and UCoGe crystals whose resistivity is presented (residual resistivity ratios of 11, 25, and 45, respectively; see also Refs. [\onlinecite{Knafo2019},\onlinecite{Knafo2012}]).  In these systems, due to a sensibility of the physical properties to the field direction, a strong crystal mosaicity can lead to broader anomalies in the field variations of $A$ and $M$. However, it is difficult to infer whether the quality of the URhSi crystals is related - or not - with the broad nature of the pseudo-metamagnetic crossover reported here. Further investigations are needed to clarify this point.

A comparison of the magnetic-field-temperature phase diagrams of URhSi, URhGe, UTe$_2$, and UCoGe in a field $\mathbf{H}\parallel\mathbf{b}$ is shown in Figure \ref{Fig6}(a) (see also Ref. [\onlinecite{Knafo2012}]). For all samples, the metamagnetic field decreases with increasing temperature, before vanishing in temperatures higher than $T_C$ in URhGe and URhSi. In UCoGe, the situation is different since anomalies are observed at the metamagnetic field under temperatures up to 20~K, i.e., much higher than $T_C = 3$~K, and seem to disappear at temperatures higher than $T_{\chi}^{max}=38$~K. Interestingly, in URhGe the magnetic susceptibility measured for $\mathbf{H}\parallel\mathbf{b}$ is maximal at $T_C=T_{\chi}^{max}=9.5$~K and this borderline is connected with the metamagnetic field $H_m$. In UTe$_2$, which is paramagnetic, contrary to the three other samples, a sharp first-order transition is observed at $H_m$ at low temperature and ends into a critical endpoint at $T_{CEP}=7$~K, above which a pseudo-metamagnetic crossover is established \cite{Knafo2019}. In URhGe and UTe$_2$, the sharpness of the maximum in $\chi(T)$ at $T_{\chi}^{max}$ is related with the sharpness of the anomalies at $H_m$. Oppositely, in URhSi the broad nature of the crossover at $H_m$ may be related with the non-observation of a maximum in $\chi(T)$ (see Ref. [\onlinecite{Matsuda2019}]).

Figure \ref{Fig6} (b) shows that, similarly to the URhSi case, a maximum of $A$ at $H_m$ is observed in URhGe and UTe$_2$ under $\mathbf{H}\parallel\mathbf{b}$  [\onlinecite{Knafo2019},\onlinecite{Miyake2008}]. The maximum of $A$ is sharper and more intense in URhGe and UTe$_2$ than in URhSi, indicating stronger magnetic fluctuations in these two materials. A study of URhGe further showed that, the higher the residual resistivity ratio is, the higher the $A$ coefficient and the reentrant superconducting temperature $T_{sc}$ are.\cite{Miyake2008} The critical magnetic fluctuations responsible for the enhancement of $A$ are also suspected to drive to field-reentrant superconductivity in URhGe and UTe$_2$. On one side, the relatively small enhancement of $A$ at $H_m$ in URhSi suggests that field-induced superconductivity may not be expected at $H_m$. On the other side, its similarities with URhGe, and to a lesser extend with UCoGe and UTe$_2$, stress that URhSi could be a candidate for superconductivity, once the condition of higher-quality single crystals (i.e., with a residual resistivity ratio at least $>10$) would be fulfilled. Indeed, crystal imperfections are known to destroy unconventional superconductivity.

In heavy-fermion Ising paramagnets in a field applied along their easy magnetic axis, a correlated-paramagnetic (CPM) regime is delimited by the temperature $T_{\chi}^{max}$, where a maximum occurs in $\chi(T)$, and by the pseudo-metamagnetic field $H_m$, where a moment polarization occurs.\cite{Aoki2013} In these systems, the CPM regime is characterized by strong antiferromagnetic fluctuations and often indicates the proximity of an antiferromagnetic instability. In the case of U-based ferromagnets in a field applied along a hard magnetic axis, no microscopic picture has been proposed yet to describe the maximum in the magnetic susceptibility and its relation with metamagnetism. Interestingly, the ferromagnet URhGe is in the vicinity of an antiferromagnetic instability, which can be induced by Ir-doping on the Rh-site [\onlinecite{Pospisil2018_PRB},\onlinecite{Pospisil2017}]). Further studies are now needed to identify how the physics, including superconductivity, of the U-based ferromagnets or nearly-ferromagnets is affected by their proximity to ferromagnetic and antiferromagnetic quantum phase transitions.

\section{Conclusion}

In summary, we have shown that for $\mathbf{H}\parallel\mathbf{b}$ a pseudo-metamagnetic crossover occurs at $\mu_0H_m\simeq30-40$~T in the ferromagnet URhSi. Similar moment polarization processes have been reported in the isostructural ferromagnetic superconductors URhGe and UCoGe for $\mathbf{H}\parallel\mathbf{b}$. Its strong similarities with URhGe, UCoGe and UTe$_2$ suggest that URhSi is a potential candidate for zero-field and field-induced superconductivity, with the condition that high-quality single crystals could be grown. Future experimental and theoretical works are needed to achieve a microscopic description of the magnetic interactions, with the aim to describe the transitions and crossovers occurring at $T_C$, $T_{\chi}^{max}$ and $H_m$. Another challenge will be to determine whether the magnetic fluctuations are coupled or not to a Fermi surface instability.

\section*{Acknowledgements}

We acknowledge Atsushi Miyake for sending us his data on the magnetization of UTe$_2$ and on the $A$ coefficient of URhGe. This work at the LNCMI was supported by the Programme Investissements d'Avenir under the project ANR-11-IDEX-0002-
02 (reference ANR-10-LABX-0037-NEXT). We acknowledge the financial support of the Cross-Disciplinary Program on Instrumentation and Detection of CEA, the French Alternative Energies and Atomic Energy Commission, and KAKENHI (JP15H05882, JP15H05884, JP15K21732, JP16H04006, JP15H05745, JP19H00646).

\end{document}